\documentclass[conference]{IEEEtran}
\makeatletter
\def\ps@headings{%
\def\@oddhead{\mbox{}\scriptsize\rightmark \hfil \thepage}%
\def\@evenhead{\scriptsize\thepage \hfil \leftmark\mbox{}}%
\def\@oddfoot{}%
\def\@evenfoot{}}
\makeatother
\usepackage[dvips]{graphicx}
\usepackage{graphics}
\usepackage{subfigure}
\usepackage{amssymb}
\usepackage{amsmath}
\usepackage{amsfonts}         
\usepackage{color}            
\usepackage{epic}             
\usepackage{eepic}            
\usepackage{rotating}         
\usepackage{type1cm}          
\usepackage{epsfig}
\usepackage{amsthm}
\usepackage{amsmath}
\usepackage{graphicx}
\usepackage{graphics}
\usepackage{epsfig}
\usepackage{latexsym}
\usepackage{amsfonts}
\usepackage{paralist}
\usepackage{xspace}
\usepackage{mathrsfs}
\usepackage{psfrag}
\usepackage{color}
\usepackage[ruled]{algorithm}
\usepackage[noend]{algorithmic}

\newtheorem{theorem}{Theorem}[section]
\newtheorem{lemma}[theorem]{Lemma}

\def\participant{{\textbf{p}}}
\def\aggreg{{\mathcal{A}}}
\def\ngroup{{\mathbb{G}}}
\def\zgroup{{\mathbb{Z}}}

\begin{document}

\title{Privacy-Preserving Data Aggregation without Secure Channel:  Multivariate Polynomial Evaluation}


\author{\IEEEauthorblockN{Taeho Jung$^\mathcal{y}$,
 XuFei Mao$^\mathcal{z}$, Xiang-Yang Li$^\mathcal{y}$, Shao-Jie Tang$^\mathcal{y}$,
 Wei Gong$^\mathcal{z}$, and Lan Zhang$^\mathcal{z}$}
\IEEEauthorblockA{$^\mathcal{y}$Department of Computer Science, Illinois Institute of Technology, Chicago, IL\\
$^\mathcal{z}$School of Software, TNLIST, Tsinghua University, Beijing}
}

\maketitle

\footnotetext[1]{The research of authors is partially supported by NSFC under Grant No. 61170216, No. 61228202, and No. 61272426, China 973 Program under Grant No.2011CB302705, China Postdoctoral Science Foundation funded project under grant No. 2012M510029, NSF CNS-0832120, NSF CNS-1035894, NSF ECCS-1247944.}

\begin{abstract}
Much research has been conducted to securely outsource multiple parties' data aggregation to an untrusted aggregator without disclosing each individual's privately owned data, or to enable multiple parties to jointly aggregate their data while preserving privacy. However, those works either require  secure pair-wise communication channels or suffer from high complexity. 
In this paper, we consider how an external aggregator or multiple parties can learn some algebraic statistics (e.g., sum, product) over participants' privately owned data while preserving the data privacy. We assume all channels are subject to eavesdropping attacks, and all the communications throughout the aggregation are open to others. We propose several protocols that successfully guarantee data privacy under this weak assumption while limiting both the communication and computation complexity of each participant to a small constant.
\end{abstract} 

\IEEEpeerreviewmaketitle

\section{Introduction}

The Privacy-preserving data aggregation problem has long been a hot research issue in the field of applied cryptography. In numerous real life applications such as crowd sourcing or mobile cloud computing, individuals need to provide their sensitive data (location-related or personal-information-related) to receive specific services from the entire system (e.g., location based services or mobile based social networking services). There are usually two different models in this problem: 1) an external aggregator collects the data and wants to conduct an aggregation function on participants' data (e.g., crowd sourcing); 2) participants themselves are willing to jointly compute a specific aggregation function whose input data is co-provided by themselves (e.g., social networking services). However, the individual's data should be kept secret, and the aggregator or other participants are not supposed to learn any useful information about it. Secure Multi-party Computation (SMC), Homomorphic Encryption (HE) and other cryptographic methodologies can be partially or fully exploited to solve this problem, but they are subject to some restrictions in this problem.

Secure Multi-party Computation (SMC) was first formally introduced by Yao \cite{yao1982protocols} in 1982 as Secure Two-Party Computation. 
Generally, it enables $n$ parties who want to jointly and privately compute a function 
\begin{displaymath}
f(x_1,x_2,\cdots,x_n)=\{y_1,y_2,\cdots,y_n\}
\end{displaymath}
where $x_i$ is the input of the participant $i$, and the result $y_i$ is returned to the participant $i$ only. Each result can be relevant to all input $x_i$'s, and each participant $i$ knows nothing but his own result $y_i$. One could let the function in SMC output only one uniform result to all or parts of participants, which is the algebraic aggregation of their input data. Then the privacy-preserving data aggregation problem seems to be solved by this approach. However this actually does not completely solve our problem because interactive invocation is required for participants in synchronous SMC (e.g., \cite{goldreich1998secure}), which leads to high communication and computation complexity, which will be compared in the Section \ref{section:implementation}. Even in the asynchronous SMC, the computation complexity is still too high for practical applications.

Homomorphic Encryption (HE) allows direct addition and multiplication of ciphertexts while preserving decryptability. That is, $\text{Enc}(m_1)\otimes \text{Enc}(m_2)=\text{Enc}(m_1 \times m_2)$, where $\text{Enc}(m)$ stands for the ciphertext of $m$, and $\otimes$, $\times$ refer to the homomorphic operations on the ciphertext and plaintexts respectively. One could also try to solve our problem using this technique, but HE uses the same decryption key for original data and the aggregated data. That is, the operator who executes homomorphic operations upon the ciphertexts are not authorized to achieve the final result. This forbids aggregator from decrypting the aggregated result, because if the aggregator is allowed to decrypt the final result, he can also decrypt the individual ciphertext received, which contradicts our motivation. Also, because the size of the plaintext space is limited, the number of addition and multiplication operations executed upon ciphertexts was limited until Gentry \textit{et al.} proposed a fully homomorphic encryption scheme \cite{gentry2009fully} and implemented it in \cite{gentry2011implementing}. However, Lauter \textit{et al.} pointed out in \cite{lauter2011can} that the complexity of general HE is too high to use in real application. Lauter also proposed a HE scheme which sacrificed possible number of multiplications for speed, but it still needs too much time to execute homomorphic operations on ciphertexts. 

\begin{table}[!tH]
\centering
\begin{tabular}{c|c}
\hline\hline
\multicolumn{2}{c}{Secure Multi-party Computation}\\\hline
Pros & different outputs for different participants  \\\hline
Cons & high complexity due to the computation based on garbled circuit  \\
 & frequent interactions required for synchronous SMC\\\hline\hline
\multicolumn{2}{c}{Homomorphic Encryption}\\\hline
Pros & efficient if \# of multiplcations is restricted \\\hline
Cons & decrypter can decrypt both aggregated data and individual data \\
 & trade-off between \# of multiplications and complexity exists \\\hline\hline
\end{tabular}
\end{table}

Besides the aforementioned drawbacks, both SMC and HE require an initialization phase during which participants request keys from key issuers via secure channel. This could be a security hole since the security of those schemes relies on the assumption that keys are disclosed to authorized participants only.
In this paper, we revisit the classic privacy preserving data aggregation problem. Our goal is to design efficient protocols without relying on a trusted authority and secure pair-wise communication channels. 
The main contributions of this paper are:
\begin{compactitem}
\item \emph{Formulation of a model without secure channel}:
Different from many other models in privacy-preserving data aggregation problem, our model does not require a secure communication channel throughout the protocol. 

\item \emph{Efficient protocol in linear time}:
The total communication and computation complexity of our work is proportional  to the number of participants $n$, while the complexities of many similar works are proportional to $n^2$. We do not use complicated encryption protocols, which makes our system much faster than other proposed systems.

\item \emph{General Multivariate Polynomial Evaluation}:
We generalize the privacy-preserving data aggregation to secure multivariate polynomial evaluation whose inputs are jointly provided by multiple parties. That is, our scheme enables multiple parties to securely compute
\begin{displaymath}
f(\{x_1,\cdots,x_n\})=\sum_{k=1}^{m} {c_k(\prod_{i=1}^{n} {x_{i}^{d_{i,k}}})}
\end{displaymath}
where the data $x_i$ is a privately known data by user $i$.
\end{compactitem}

Note that our general format of data aggregation can be directly used to express various statistical values. For example,  $\sum_{i=1}^{n}x_i$ can easily be achieved while preserving privacy, and thus the mean $\mu=\sum_{i=1}^{n}x_i/n$ can be computed with privacy-preserving. Given the mean $\mu$, $n\mu^2+\sum_{i=1}^{n}{(x_i^2-2x_i\mu)}$ can be achieved from the polynomial, and this divided by $n$ is the population variance. Similarly, other statistical values are also achievable (e.g., \emph{sample skewness},\emph{$k$-th moment}, \emph{mean square weighted deviation}, \emph{regression}, and \emph{randomness} test) based on our general multi-variate polynomial.
Although our methods are proposed for computing the value of a multi-variate polynomial function where the input of each participant is assumed to be an integer, our methods can be generalized for functions (such as dot product) where the input of each participant is a vector.

The rest of the paper is organized as follows.
We present the system model and necessary background in Section \ref{sec:prelim}.
In Section \ref{sec:secure-channel},
 we  analyze the needed number of communications with secure communication channels when users communicate randomly.
We first address the privacy preserving summation and production in Section \ref{sec:simple} by presenting two efficient protocols.
Based on these protocols, we then present an efficient protocol for  general multi-variate polynomial evaluation in Section \ref{sec:general}.
In Section \ref{sec:analysis}, we present detailed analysis of the correctness, complexity, and  security of our protocols.
Performance evaluation of our protocols is reported in Section \ref{sec:evalaution}.
We compare our protocol with the ones based on SMC or HE.
We then conclude the paper with the discussion of some future work in Section \ref{sec:conclusion}. 

\section{Related Work}

Many novel protocols have been proposed for privacy-preserving data aggregation or in general secure multi-party computation. 
Castelluccia \textit{et al.} \cite{castelluccia2009efficient} presented a provable secure and efficient aggregation of encrypted data in WSN, which is extended from \cite{castelluccia2005efficient}. They designed a symmetric key homomorphic encryption scheme which is additively homomorphic to conduct the aggregation operations on the ciphertexts. Their scheme uses modular addition, so the scheme is good for CPU-bounded devices such as sensor nodes in WSN. Their scheme can also efficiently compute various statistical values such as mean, variance and deviation. However, since they used the symmetric homomorphic encryption, their aggregator could decrypt each individual sensor's data, and they assumed the trusted aggregator in their model.

Sheikh \textit{et al.} \cite{sheikh2009privacy} proposed a $k$-secure sum protocol, which is motivated by the work of Clifton \textit{et al.} \cite{clifton2002tools}. They significantly reduced the probability of data leakage in \cite{clifton2002tools} by segmenting the data block of individual party, and distributing segments to other parties. Here, sum of each party's segments is his data, therefore the final sum of all segments are sum of all parties' data. This scheme can be easily converted to $k$-secure product protocol by converting each addition to multiplication. Similar to our protocol, one can combine their sum protocol and converted product protocol to achieve a privacy-preserving multivariate polynomial evaluation protocol. However, pair-wise unique secure communication channels should be given between each pair of users such that only the receiver and the sender know the transmitted segment. Otherwise, each party's secret data can be calculated by performing $O(k)$ computations. In this paper, we remove the limitation of using secure communication channels.

The work of He \textit{et al.} \cite{he2007pda} is similar to Sheikh \textit{et al.}'s work. They proposed two privacy-preserving data aggregation schemes for wireless sensor networks: the Cluster-Based Private Data Aggregation (CPDA) and the Slice-Mix-AggRegaTe (SMART). In CPDA, sensor nodes form clusters randomly and collectively compute the aggregate result within each cluster. In the improved SMART, each node segments its data into $n$ slices and distributes $n-1$ slices to nearest nodes via secure channel. However, they only supports additions, and since each data is segmented, communication overhead per node is linear to the number of slices $n$.

Shi \textit{et al.} \cite{shi2011privacy} proposed a construction that $n$ participants periodically upload encrypted values to an aggregator, and the aggregator computes the sum of those values without learning anything else. This scheme is close to our solution to the multivariate polynomial evaluation problem, but they assumed a trusted key dealer in their model. The key dealer distributes random key $k_i$ to participant $i$ and key $k_0$ to the aggregator, where  $\Pi_{i=0}^{n} k_i =1$, and the ciphertext is in the format of $C_i=k_i\cdot g^{x_i}$. Here, $g$ is a generator, $k_i$ is a participant's key and $x_i$ is his data (for $i=1,2, \cdots n$). Then, the aggregator can recover the sum $\sum_{i=1}^n x_i$ iff he received ciphertexts from all of the participants. He computes $k_0 \Pi_{i=1}^n C_i$ to get $g^{\sum_{i=1}^n x_i}$, and uses brute-force search to find the $\sum_{i=1}^n x_i$ or uses Pollard's lambda method \cite{menezes1997handbook} to calculate it. This kind of brute-force decryption limits the space of plaintext due to the hardness of the discrete logarithm problem, otherwise no deterministic algorithm can decrypt their ciphertext in polynomial time. The security of their scheme relies on the security of keys $k_i$.

In our scheme, the trusted aggregator in \cite{castelluccia2009efficient}\cite{castelluccia2005efficient} is removed since data privacy against the aggregator is also a top concern these days. Unlike \cite{he2007pda}\cite{sheikh2009privacy}, we assumed insecure channels, which enabled us to get rid of  expensive and vulnerable key pre-distribution. We  did not segment each individual's data, our protocols only incur constant  communication overhead for each participant. Our scheme is also based on the hardness of the discrete logarithm problem like \cite{shi2011privacy}, but we do not trivially employ brute-force manner in decryption, instead, we employ our novel efficient protocols for sum and product calculation. 

\section{System Models and Preliminary}
\label{sec:prelim}

\subsection{System Model and Problem Definition}

Assume that there are $n$ participants $\{\participant_1, \participant_2, \cdots, \participant_n\}$, and each participant $\participant_i$ has a privately known data $x_i$ from a group $\ngroup_1$.
The privacy-preserving data aggregation problem (or secure multivariate polynomial evaluation problem) is to compute some multivariate polynomial of $x_i$ jointly or by an aggregator while preserving the data privacy.
Assume that there is a group of $m$ powers $\{d_{i,k}\in\zgroup_{q} \mid k=1, 2,\cdots,m \}$ for each $\participant_i$ and $m$ coefficients $\{c_{k}  \mid k=1,\cdots,m, c_k \in \ngroup_1 \}$.
The objective of the aggregator or the participants is to  compute the following polynomial without knowing any individual $x_i$:
\begin{align}
f(\textbf{x})=\sum_{k=1}^{m}(c_{k} \prod_{i=1}^{n} {x^{d_{i,k}}_i}) 
\label{equation:polynomial}
\end{align}
Here vector $\textbf{x}=(x_1,x_2, \cdots, x_n)$.
For simplicity, we assume that the final result $f(\textbf{x})$ is positive and bounded from above by a large prime number $P$.
We assume all of the powers $d_{i,k}$'s and coefficients $c_{k}$'s are open to any participant as well as the attackers. This is a natural assumption since the powers and coefficients uniquely determine a multivariate polynomial, and the polynomial is supposed to be public.

We employ two different models in this paper: \emph{One Aggregator Model} and \emph{Participants Only Model}. These two models are general cases we are faced with in real applications.

\textbf{One Aggregator Model}:
In the first model, we have one aggregator $\aggreg$ who wants to compute the function $f(\textbf{x})$. We assume the aggregator is untrustful and curious. That is, he always eavesdrops the communications between participants and tries to harvest their input data. We also assume participants do not trust each other and that they are curious as well, however, they will follow the protocol in general. We could also consider having multiple aggregators, but this is a simple extension which can be trivially achieved from our first model. We call this model the \emph{One Aggregator Model}.
Note that in this model, any single participant $\participant_i$ is not allowed to compute the final result $f(\textbf{x})$.

\textbf{Participants Only Model}:
The second model is similar to the first one except that there are $n$ participants only and there is no aggregator. In this model,  all the participants are equal and they all will calculate the final aggregation result $f(\textbf{x})$. We call this model the \emph{Participants Only Model}.

In both models, participants are assumed not to collude with each other. Relaxing this assumption is one of our future work.

\subsection{Additional Assumptions}

We assume that all the communication channels in our protocol are insecure. Anyone can eavesdrop them to intercept the data being transferred. 
To address the challenges of insecure communication channel, we assume that the discrete logarithm problem
is computationally hard if: 1) the orders of the integer groups are large prime numbers; 2) the involved integer numbers are large numbers. The security of our scheme relies on this assumption.
We further assume that there is a secure pseudorandom function (PRF) which can choose a random element from a group such that this element is computationally indistinguishable to uniform random.

We also assume that user authentication was in place to authenticate each participants if needed.
 We note that Dong \emph{et al.} \cite{dong2011secure} investigated verifiable privacy-preserving dot production of two vectors and Zhang \textit{et al.} \cite{zhang2013verifiable} proposed verifiable multi-party computation, both of which can be partially or fully exploited later.
 Designing privacy preserving data aggregation while providing  verification of the correctness of the provided data is a future work.

\subsection{Discrete Logarithm Problem}
Let $\ngroup\subset\zgroup_p$ be a cyclic multiplicative integer group, where $p$ is a large prime number, and $g$ be a generator of it. Then, for all $h\in \ngroup$, $h$ can be written as $h=g^k$ for some integer $k$, and any integers are congruent modulo $p$. The discrete logarithm problem is defined as follows: given an element $h\in \ngroup$, find the integer $k$ such that $g^b=h$.

The famous Decision Diffie-Hellman (DDH) problem proposed by Diffie and Hellman in \cite{diffie1976new} is derived from this assumption. DDH problem is widely exploited in the field of cryptography (e.g., El Gamal encryption \cite{elgamal1985public} and other cryptographic security protocols such as CP-ABE \cite{bethencourt2007ciphertext}) as discussed in \cite{boneh1998decision}.
Our protocol is based on the  assumption that it is computational expensive to solve the discrete logarithm problem as in other similar research works (\cite{jung2013cloud,li2013search,zhang2013verifiable}).

\section{Achieving sum  Under Secured Communication Channel}
\label{sec:secure-channel}

Before introducing our aggregation scheme without secure communication channel, we first describe the basic idea of randomized secure sum calculation under secured communication channel (It can be trivially converted to secure product calculation). The basic idea came from Clifton \textit{et al.} \cite{clifton2002tools}, which is also reviewed in \cite{verykios2004state}, but we found their setting imposed unnecessary communication overhead, and we reduced it while maintaining the same security level. Assume participants $\participant_1, \participant_2,\cdots, \participant_n$ are arranged in a ring for computation purpose. Each participant $\participant_i$ itself breaks its privately owned data block $x_i$  into $k$ segments $s_{i,j}$ such that the sum of all $k$ segments is equal to the value of the data block. The value of each segment is randomly decided. For sum, we can simply assign random values to segments $s_{i,j}$ ($1 \le j \le k-1$) and let  $s_{i,k}=x_i - \sum_{j = 1}^{k-1} s_{i,j}$. Similar method can be used for product. In this scheme, each participant randomly selects $k-1$ participants and transmit  each of those participants a distinctive segment $s_{i,j}$. Thus at the end of this redistribution each of participants holds several segments within which one segment belongs to itself and the rest belongs to some other participants. The receiving participant adds all its received segments and transmits its result to the next participant in the ring. This process is repeated until all the segments of all the participants are added and the sum is announced by the aggregator.

Recall that there are $n$ participants and each participant randomly selects $k-1$ participants to distribute its segments. Clearly, a larger $k$ provides better computation privacy, however it also causes larger communication overhead which is not desirable. In the rest of this section, we are interested at finding an appropriate $k$ in order to reduce the communication cost while preserving computation privacy.

In particular, we aim at selecting the smallest $k$ to ensure that each participant holds at least one segment from the other participants after redistribution.
We can view this problem as placing identical and indistinguishable balls into $n$ distinguishable
(numbered) bins. This problem has been extensively studied and well-understood and the following lemma can be proved by simple union bound:

\begin{lemma}
Let $\epsilon \in (0,1)$ be a constant. If we randomly place $(1 + \epsilon)n \ln n$ balls into
$n$ bins, with probability at least $1-\frac{1}{n^\epsilon}$, all the $n$ bins are filled.
\end{lemma}

Assume that each participant will \emph{randomly} select $k-1$ participants (including itself) for redistribution. By treating each round of redistribution as one trial in coupon's collector problem, we are able to prove that each participant only needs to redistribute $((1 + \epsilon)n \ln n)/n = (1 + \epsilon)\ln n$ segments to other participants to ensure that every participant receives at least one segment with high probability.
However, different from previous assumption, each participant will select $k-1$ participants except itself to redistribute its segments in our scheme. Therefore, we need one more round redistribution for each participant to ensure that every participant will receive at least one copy from other participants with high probability.

\begin{theorem}
Let $\epsilon \in (0,1)$ be a constant. If each participant randomly selects $(1 + \epsilon)\ln n + 1$ participants to redistribute its segments, with probability at least $1-\frac{1}{n^\epsilon}$, each participant receives at least one segment from the other participants.
\end{theorem}

This theorem reveals that by setting $k$ to the order of $\ln n$, we are able to preserve the computation privacy. Compared with traditional secure sum protocol, our scheme dramatically reduce the communication complexity. However, we assume that the communication channel among participants are secure in above scheme. In the rest of this paper, we try to tackle the secure aggregation problem under unsecured channels.

\section{Efficient Protocols for Sum and Product}\label{section:preliminaries}
\label{sec:simple}

In this section, we present two novel calculation protocols for each model which preserve individual's data privacy. These four protocols will serve as bases of our solution to privacy-preserving data aggregation problem. For simplicity, we assume all coefficients $c_k$ ($k \in[1, m]$) and powers $d_{i,k}$ ($i\in[1,n], k\in[1,m]$) of the polynomial $f(\textbf{x})=\sum_{k=1}^m c_k{(\prod_{i=1}^n {x_i^{d_{i,k}}})}$ are known to every participant $\participant_i$.
Table \ref{tab:notions} summarizes the main notations used in this paper.

\begin{table}[!h]
\caption{Notations of symbols used in our protocols}
\label{tab:notions}
\centering
\begin{tabular}{l|l}
\hline\hline
$\participant_i$ & $i$-th participant in data aggregation\\
$\aggreg$ & Aggregator \\
$\ngroup_1,\ngroup_2$ & multiplicative cyclic integer groups \\
$g_1,g_2$ & generators of above groups \\
$d_{i,k}$ & power of $x_i^{d_{i,k}}$ \\
$c_k$ & coefficient of $c_k\sum_{i=1}^{n}{x_i^{d_{i,k}}}$ \\
$r_i,\hat{r}_i$ & randomly chosen numbers \\
\hline\hline
\end{tabular}
\end{table}

\subsection{Product Protocol - Participants Only Model}

Firstly, we assume that all participants together
 want to compute the value $f(\textbf{x})=\prod_{i} x_i$ given their privately known values $x_i \in \zgroup_{p}$.
The basic idea of our protocol is to find some random integers $R_i \in \zgroup_{p}$ such that
$\prod_{i} R_i =1 \mod p$ and the user $\participant_i$ can compute the random number $R_i$ easily while it is computationally expensive for other participants to compute the value $R_i$.

Let $\ngroup_1\subset \zgroup_p$ be a cyclic multiplicative group of prime order $p$ and $g_1$ be its generator. Then our protocol for privacy preserving production $\Pi_i x_i$ has the following steps:
\textsf{Setup}, \textsf{Encrypt}, \textsf{Product}.

\medskip
\textsf{Setup} $\rightarrow$ $r_i\in\zgroup_q, R_i=(g_1^{r_{i+1}}/g_1^{r_{i-1}})^{r_i}\in\ngroup_1$

We assume all participants are arranged in a ring for computation purpose. The ring can be formed according to the lexicographical order of the MAC address or even the geographical location. It is out of our scope to consider this problem.  Each $\participant_i (i\in\{1,\cdots,n\})$ randomly chooses a secret integer $r_i\in\zgroup_q$ using PRF and calculates a public parameter $g_1^{r_i}\in\ngroup_1$. Then, each $\participant_i$ shares $Y_i=g_1^{r_i} \mod p$ with $\participant_{i-1}$ and $\participant_{i+1}$ (here $\participant_{n+1}=\participant_1$ and $\participant_0=\participant_{n}$). 

After a round of exchanges, the participant $\participant_i$ computes the number $R_i=(Y_{i+1}/Y_{i-1})^{r_i}=(g_1^{r_{i+1}}/g_1^{r_{i-1}})^{r_i} \mod p$ and keeps this number $R_i$ secret. Note $\participant_1$ calculates $(g_1^{r_{2}}/g_1^{r_{n}})^{r_1}$ and $\participant_n$ calculates $(g_1^{r_{1}}/g_1^{r_{(n-1)}})^{r_n}$.

\begin{figure}[!h]
\begin{center}
\begin{tabular}{cc}
\scalebox{0.53}{\input{cycle-new.pstex_t}}  \ \ &
\scalebox{0.53}{\input{cycle3-new.pstex_t}}\\
(a)Participants only model & (b) One aggregator model
\end{tabular}
\caption{Communications in \textsf{Setup}}
\label{fig:cycle}\label{fig:cycle3}
\end{center}
\end{figure}

\textsf{Encrypt($x_i$)} $\rightarrow$ $C_{i}\in\ngroup_1$

When a product is needed, every $\participant_i$ creates the ciphertext:
\begin{displaymath}
C_{i}:=x_i\cdot R_i =x_i \cdot (g_1^{r_{i+1}}/g_1^{r_{i-1}})^{r_i} \mod p
\end{displaymath}
where $x_i$ is his private input data. If he does not want to participate in the multiplication, he can simply set $x_i:=1$. Then, he broadcasts this ciphertext.
\medskip
\begin{figure}[!th]
\begin{center}
\begin{tabular}{cc}
\scalebox{0.61}{\input{enc_participant-new.pstex_t}}  \ \ &
\scalebox{0.61}{\input{enc_oneaggr-new.pstex_t}}\\
(a)Participants only model & (b) One aggregator model
\end{tabular}
\caption{Communications in \textsf{Encrypt}}
\label{fig:enc_participant}\label{fig:enc_oneaggre}
\end{center}
\end{figure}

\textsf{Product($\{C_{1},C_{2},\cdots,C_{n}\}$)} $\rightarrow$ $\prod_{i=1}^{n}{x_i}\in\ngroup_1$

Any $\participant_i$, after receiving $n$ ciphertexts $\{C_{1},C_{2},\cdots,C_{n}\}$ from all of the $\participant_i$'s, calculates the following product:
\begin{displaymath}
\prod_{i=1}^{n}{C_{i}}=\prod\limits_{i=1}^{n}{x_i} \mod p
\end{displaymath}

To make sure that we can get a correct result $\prod_{i=1}^n x_i$ without modular, we can choose $p$ to be large enough, say $p \ge M^n$, where $M$ is a known upper bound on $x_i$.

\subsection{Product Protocol - One Aggregator Model}

We use the same group used in Participants Only Model. Everything is same as the protocol above, except that the aggregator $\aggreg$ acts as the $(n+1)$-th participant $\participant_{n+1}$.
In other words, there are $n+1$ ``participants'' now.
The second difference is that, each participant $\participant_i$ will send the ciphertext $C_i$ to the aggregator, instead of broadcasting to all participants.
The aggregator $\aggreg$ will \emph{not} announce its random number $R_{n+1}=(g_1^{r_{1}}/g_1^{r_{n}})^{r_{n+1}}$ to any regular participants.

Each participant $\participant_i$, $i \in [1,n]$, sends the ciphertext $C_i=R_i \cdot x_i$ to the aggregator $\aggreg$.
 The aggregator  $\aggreg$ then calculates
\begin{displaymath}
(g_1^{r_{1}}/g_1^{r_{n}})^{r_{n+1}}\prod_{i=1}^{n}{x_{i}}=\prod_{i=1}^{n}{x_i}
\end{displaymath}
to achieve the final product, where $r_{n+1}$ is the random number generated by $\aggreg$.

\subsection{Sum Protocol - Participants Only Model}\label{section:sum}

Here we assume that all participants together
 want to compute the value $f(\textbf{x})=\sum_{i=1}^n x_i$ given their privately known values $x_i \in \zgroup_{p}$.
 It seems that we can still exploit the method used for computing product by finding random numbers $R_i$ such that $\sum_{i=1}^n R_i =0$. We found that it is challenging to find such a number $R_i$ while preserve privacy and security.
The basic idea of our protocol is to convert the sum of numbers into production of numbers.
Previous solution \cite{shi2011privacy} essentially applied this approach also by computing the product of $\prod_{i=1}^n g^{x_i}=g^{\sum_{i=1}^n x_i}$.
Then find $\sum_{i=1}^n x_i$ by computing the discrete logarithm of the product.
As discrete logarithm is computational expensive, we will not adopt this method.
 Instead, we propose a computational efficient method here.

In a nutshell, we exploit the modular property below to achieve the privacy preserving sum protocol.
\begin{align}
(1+p)^m = \sum_{i=0}^{m}{{m\choose i}p^i} = 1+mp \mod p^2
\label{equation:property}
\end{align}
From the Equation (\ref{equation:property}), we conclude that
\begin{displaymath}
\prod_{i=1}^n {(1+p)^{x_i}} = \prod_{i=1}^n (1+p \cdot x_i)=(1+p\sum_i{x_i}) \mod p^2.
\end{displaymath}

Our protocol works as follows. Let $\ngroup_2\subset \zgroup_{p^2}$ be a cyclic multiplicative group of order $p(p-1)$ and $g_2$ be its generator, where $p$ is a prime number. Then our protocol for privacy preserving summation $\Pi_i x_i$ has the following steps:
\textsf{Setup}, \textsf{Encrypt}, \textsf{Sum}.

\medskip
\textsf{Setup} $\rightarrow$ $r_i\in\zgroup_{pq}, R_i=(g_2^{r_{i+1}}/g_2^{r_{i-1}})^{r_i}$

Remember that participants are arranged in a circle. $\participant_i$ uses $PRF$ to randomly pick a secret number $r_i\in\zgroup_{pq}$, and calculates a public parameter $g_2^{r_i}$. Then, he shares $g_2^{r_i}$ with $\participant_{i+1}$ and $\participant_{i-1}$. Similar to the product calculation protocol, $\participant_n$ shares his public parameter with his $\participant_{(n-1)}$ and $\participant_1$, and $\participant_1$ shares his public parameter with $\participant_{2}$ and $\participant_n$.

After a round of exchanges, each $\participant_i$ calculates $R_i=(g_2^{r_{i+1}}/g_2^{r_{i-1}})^{r_i}$ and keeps this secret.

\medskip
\textsf{Encrypt($x_i, R_i$)} $\rightarrow$ $C_i\in\ngroup_2$

This algorithm crosses over two different integer groups: $\ngroup_1$ and $\ngroup_2$. Each $\participant_i$ first calculates $(1+x_i\cdot p) \mod p^2$. Note that $x_i\in\ngroup_1$, and it is temporarily treated as an element in $\ngroup_2$, but this does not affect the last value of the result since operations in $\ngroup_2$ are modulo $p^2$. Then, he multiplies the secret parameter $R_i=(g_2^{r_{i+1}}/g_2^{r_{i-1}})^{r_i}$ to it to get the ciphertext:
\begin{displaymath}
C_i=(1+x_i\cdot p) \cdot R_i
\end{displaymath}

After all, each participant broadcasts his ciphertext to each others.

\textsf{Sum($\{C_1,C_2,\cdots,C_n\}$)} $\rightarrow$ $\sum_{k=1}^{n}{x_i}\in\ngroup_1$.

Each participant, after receiving the ciphertexts from all of other participants, calculates the following $C\in\ngroup_2$:
\begin{displaymath}
C=\prod_{i=1}^{n} C_i=(1+p\sum_{i=1}^{n}{x_i}) \mod p^2
\end{displaymath}
Then, he calculates
$(C-1)/p \mod p= \sum_{i=1}^{n}{x_i} \mod p$
 to recover the final sum.

\subsection{Sum Protocol - One Aggregator Model}

Similar to the product protocol for One Aggregator Model, everything is the same except that $\aggreg$ acts as ($n+1$)-th participant in this model. The participants send their ciphertexts to $\aggreg$, and $\aggreg$ calculates
\[C=(g_2^{r_{1}}/g_2^{r_{n}})^{r_{n+1}}\prod_{i=1}^{n} C_i
=(1+p\sum_{i=1}^{n}{x_i}) \mod p^2
\]
Then, he can compute the final sum result $\sum_{i=1}^{n}{x_i}$.

\section{Efficient Protocols for General Multivariate Polynomial}
\label{sec:general}

Now we are ready to present our efficient privacy preserving protocols for evaluating a multivariate polynomials.
Our protocol is based on the efficient protocols for sum and production presented in the previous section.

\subsection{One Aggregator Model}\label{section:model1}
The calculation of the polynomial \ref{equation:polynomial} can be divided into $nm$ multiplications and $m$ additions. In this section we show how to conduct a joint calculation of $m$ products and one sum while preserving individual's data privacy in the One Aggregator Model. Different from the protocols in the Section \ref{section:preliminaries}, those broadcast ciphertexts are not broadcast this time, they are sent to the aggregator instead. The purpose of this small change is only for reducing communication complexity, and from the security perspective, this is just same as broadcasting since our communication channels are insecure.
\subsubsection{Basic Scheme}
All the participants execute \textsf{Setup} to initiate the system. Then, for each $k$, all the participants need to calculate $x_i^{d_{i,k}}$'s first, where $d_{i,k}$'s are powers specified by the aggregator $\aggreg$, and run the aforementioned product protocol for each $k\in [1,m]$. If $\aggreg$ does not need the data from some participant $\participant_i$, $\aggreg$ can set his powers to be 0, and if $\participant_i$ does not want to participate in the aggregation, he can simply set his input as 1.

Then, the aggregator is able to calculate $\sum_{k=1}^{m}{(c_k\prod_{i=1}^{n}{x_i^{d_{i,k}}})}$.

\subsubsection{Advanced Scheme}\label{section:securescheme}
The above \textit{Basic Scheme} preserves data privacy in our problem as long as there are at least two $x_i^{d_{i,k}}$'s not equal 1 in each following set $\{x_1^{d_{1,k}},x_2^{d_{2,k}}\cdots,x_n^{d_{n,k}}\}_{k\in\{1,\cdots,m\}}$, which will be further discussed in the Section \ref{section:special_prod}. Therefore, we exploit the aforementioned sum protocol to achieve \textit{Secure Scheme}.

All the participants execute \textsf{Setup}. Then, when executing the \textsf{Encrypt} of the product protocol, each participant checks whether his input is the only one not equal to 1 for each product $\prod_{i=1}^{n}{x_i^{d_{i,l}}}$ (i.e., his $d_{i,l}$ is the only one not equal to 0 in \{$d_{1,l}, d_{2,l},\cdots,d_{n,l}$\}). If it is, the product equals to his input data, which will directly disclose his data, so he skips it. The elements that are omitted form a set $D_{sum}=\{x_i^{d_{i,k}}\}_{k\in I_{sum}}$, where $I_{sum}$ is the set of indices $k$'s corresponding to the elements in $D_{sum}$. For each $x_i^{d_{i,k}}\in D_{sum}$, find his owner $\participant_i$ and add him into the set $P_{sum}$. There can be duplicate $\participant_i$'s in the set $P_{sum}$. The $\participant_i$'s in $P_{sum}$ need to calculate the following without knowing each other's input:
\begin{displaymath}
\sum_{\participant_i\in P_{sum}}{c_kx_i^{d_{i,k}}}
\end{displaymath}

They are called sum participants, and we assume they are ordered by non-decreasing order of their indices in $P_{sum}$ and arranged in a circle. In what follows, we denote $\participant_i$'s successor and predecessor in the $P_{sum}$ as $\participant_{i,suc}$ and $\participant_{i,pre}$ respectively. These sum participants run the sum protocol to encrypt their data and sends to the aggregator $\aggreg$.

$\aggreg$, after receiving all the sum ciphertexts, is able to calculate $\sum_{k\in I_{sum}}{c_kx_i^{d_{i,k}}}$. Then, he is able to calculate $\sum_{k=1}^{m}{(c_k\prod_{i=1}^{n}{x_i^{d_{i,k}}})}$.

\subsection{Participants Only Model}\label{section:model2}
From the One Aggregator Model, we know the combination of two protocols (product protocol and second sum protocol) proposed in Section \ref{section:preliminaries} gives the best scheme. Therefore we only show the scheme which employs both product and sum protocols.

\textit{Advanced Scheme:} Every participant executes \textsf{Setup}, and when he executes the \textsf{Encrypt} of the product protocol, he conducts the same examination as in the Section \ref{section:securescheme} above. Then, the sum participants run the sum protocol to share their sum with each other. Finally, all participants are able to calculate $\sum_{k=1}^{m}{(c_k\prod_{i=1}^{n}{x_i^{d_{i,k}}})}$ based on others' ciphertexts. 

\section{Correctness, Complexity and Security Analysis}
\label{sec:analysis}

Here we provide rigorous correctness proofs, complexity and security analysis of the protocols presented in this paper.
We also discuss when our protocols could leak information about the privately known data $x_i$ and provide
methods to address this when possible.

\subsection{Correctness}
\label{section:correctness}
Next we show the correctness of the product protocol in Section \ref{section:preliminaries}.

\subsubsection{Product Protocol}
We only provide the analysis for Participants Only model, but the correctness in One Aggregator model is easily derivable from it. After participants receive $\{C_{1},\cdots,C_{n}\}$ they conduct the following calculation:

\begin{displaymath}
\begin{split}
\prod_{i=1}^{n}{C_{i}} &=\prod_{i=1}^{n}{(x_i(g_1^{r_{i+1}}/g_1^{r_{i-1}})^{r_i})}\\
&=(\prod_{i=1}^{n}{x_i}) \prod_{i=1}^{n}{((g_1^{r_{i+1}}/g_1^{r_{i-1}})^{r_i}})\\
&=(\prod_{i=1}^{n}{x_i}) g_1^{\sum_{i=1}^{n}{(r_{i+1}r_{i}-r_{i}r_{i-1})}}
=\prod_{i=1}^{n}{x_i}
\end{split}
\end{displaymath}

Here $r_{n+1}=r_1, r_0=r_n$. Thus, the products are correctly calculated.

\subsubsection{Sum Protocol}

Similar to above, we only discuss the correctness for Participants Only Model. After participants receive $\{C_{1},\cdots,C_{n}\}$, they conduct the following calculation:
\begin{displaymath}
\begin{split}
C&=\prod_{i=1}^{n} C_i
=\prod_{i=1}^{n}{(1+x_ip)(g_2^{r_{i+1}}/g_2^{r_{i-1}})^{r_i}}\\
&=(1+p\sum_{i=1}^{n}{x_i})g_2^{\sum_{i=1}^{n}{r_{i+1}r_i-r_ir_{i-1}}}\\
&=(1+p\sum_{i=1}^{n}{x_i}) \mod p^2
\end{split}
\end{displaymath}
Thus, $(C-1)/p$ mod $p$ is indeed equal to $\sum_{i=1}^{n}{x_i}$ mod $p$.

\subsection{Security}
We discuss the security of the schemes in both One Aggregator Model and Participants Only Model in this section.

\subsubsection{Special Case of Products Calculation}\label{section:special_prod}

As mentioned in the Section \ref{section:securescheme}, if there is only one ciphertext $d_{i,k}$ is not equal to 0 in any set $\{d_{1,k},d_{2,k}\cdots,d_{n,k}\}_{k\in\{1,\cdots,m\}}$ during the products calculation, the individual data $x_i$ can be disclosed to others. This is because: \\(suppose that only $d_{i,k}$ is the only ciphertext not equal to 1 in the set $\{d_{1,k},d_{2,k}\cdots,d_{n,k}\}$)
\begin{displaymath}
\textsf{Decrypt($\{C_{1,k},C_{2,k}\cdots,C_{n,k}\}$)}=x_i
\end{displaymath}
and $x_i$ is disclosed to others if $c_k\neq 0$. Therefore, in this case, the participants should conduct additional secure sum calculation before sending the ciphertexts to others.

\subsubsection{Randomness and Group Selection}\label{section:randomnessproduct}

In fact, in the product calculation protocol, the group $\ngroup_1$ should be carefully selected to make the input $x_i$ indistinguishable to a random element. We select a cyclic multiplicative group $\ngroup_1\subset\zgroup_p$ of prime order $q$ as follows. Find two large prime numbers $p,q$ such that $p=kq+1$ for some integer $k$. Then, find a generator $h$ for $\zgroup_p$,  and set $g_1:=h^{(p-1)/q}$ modulo $p$ (clearly $g_1\neq 1$ modulo $p$). Then group $\ngroup_1$ is generated by $g_1$, whose order is $q$. Here the powers of the numbers in $\ngroup_1$ belong to an integer group $\zgroup_q$. 

Next, we show that any input data $x_i$ is computationally indistinguishable to any random element chosen from $\zgroup_p$ via PRF.

For any $i$, we have
\[C_{i}=x_i(g_1^{r_{i+1}}/g^{r_{i-1}})^{r_i} =x_ig_1^{(r_{i+1}-r_{i-1})r_i}.\]
Let $x_i$ be $g_1^{\chi_i}$ and $r_{i+1}-r_{i-1}$ be $\gamma_i$, where $\chi_i\in\zgroup_q$ and $\gamma_i\in\zgroup_q$ (This is possible since $g_1$ is a generator of the group $\ngroup_1$). Then, $C_{i}=g_1^{\chi_i} g_1^{\gamma_i r_i}$.

\begin{theorem}\label{theorem:random}
$\forall x_i, r_i\in\zgroup_q$,  $\exists \hat{r_i}, \hat{\chi}_i\in\zgroup_q$  such that
 \[g_1^{\chi_i} g_1^{\gamma_i r_i}=g_1^{\hat{\chi}_i}g_1^{\gamma_i \hat{r_i}} \mod p.\]
\end{theorem}
\begin{proof}
For any $r_i, \hat{r_i}\in\zgroup_q$, there exists $\hat{\chi}_i\in\zgroup_q$ such that:
\begin{displaymath}
\gamma_i(r_i- \hat{r_i})=\hat{\chi}_i-\chi_i \mod q
\end{displaymath}
because $q$ and $(r_i- \hat{r_i})$ are relatively prime ($q$ is a prime number). Then we have $\hat{\chi}_i\in\zgroup_q$ for any $r_i\in\zgroup_q$ such that:
\begin{displaymath}
g_1^{\gamma_i(r_i-\hat{r_i})}=g_1^{\hat{\chi}_i-\chi_i} \mod p
\Rightarrow g_1^{\chi_i} g_1^{\gamma_i r_i} =g_1^{\hat{\chi}_i} g_1^{\gamma_i  \hat{r_i}} \mod p
\end{displaymath}
This implies that given the ciphertext $C_i$, any value $x_i$ is a possible valid data that can produce this 'ciphertext' $C_i$.
\end{proof}

According to the Theorem \ref{theorem:random}, we can deduce that $\chi_i$ has the same level of randomness as $r_i$. Therefore, $g_1^{\chi_i}$ is indistinguishable to a random element in $\ngroup_1$ from other participants' or attackers' perspective, which implies

\begin{theorem}
The input $x_i$ is computationally indistinguishable to a random element chosen from $\ngroup_1$.
\end{theorem}

\subsubsection{Closure and Group Selection}

We need to guarantee that all the multiplications in the sum protocol are closed in $\ngroup_2$. Since $(1+x_ip)\cdot (g_2^{r+i}/g_2^{r-1})^{r_i}$ is the only multiplication throughout the sum protocol, we must carefully choose the group $\ngroup_2$ such that $1+x_ip\in\ngroup_2$. We let $\ngroup_2\subset \zgroup_{p^2}$ be a cyclic multiplicative group generated by $h$, which is the generator of $\zgroup_p$. Then, the order of $\ngroup_2$ is $p(p-1)$, and the powers of the numbers in $\ngroup_2$ belong to an integer group $\zgroup_{p(p-1)}$. Since $\ngroup_2 = \zgroup_{p^2} - \{x| x = k\cdot p, \text{for some integer }k\}$ and $\forall k: 1+x_ip\neq kp$, $1+x_ip$ belongs to the group $\ngroup_2$.

\subsubsection{Restriction of the Product and Sum Protocol}
In both protocols, we require that number of participants is at least 3 in Participants Only Model and at least 2 in One Aggregator Model. In Participants Only Model, if there are only 2 participants, privacy is not preservable since it is impossible to let $\participant_1$ know $x_1+x_2$ or $x_1x_2$ without knowing $x_2$. However, in One Aggregator Model, since only the aggregator $\aggreg$ knows the final result, as long as there are two participants, $\aggreg$ is not able to infer any individual's input data.

\subsection{Complexity}
We discuss the computation and communication complexity of the \textit{Advanced Scheme} for each model in this section.

\subsubsection{One Aggregator Model}
It is easy to see that the computation complexities of \textsf{Setup}, \textsf{Encrypt} and \textsf{Product} of the product protocol are $O(1)$, $O(1)$ and $O(n)$ respectively. Also, \textsf{Encrypt} is executed for $m$ times by each participant and \textsf{Product} is executed for $m$ times by the aggregator in the \textit{Advanced Scheme}.

Every participant and the aggregator exchanges $g^{r_i}$'s with each adjacent neighbor in the ring, which incurs communication of $O(|p|)$ bits in \textsf{Setup}, where $|p|$ represents the bit length of $p$. In \textsf{Encrypt}, each participant sends $m$ ciphertexts $c_k\prod_{i=1}^{n}{x_i^{d_{i,k}}}$'s to the aggregator, so the communication overhead of \textsf{Encrypt} is $O(m|p|)$ bits. Since $n$ participants are sending the ciphertexts to the aggregator, the aggregator's communication overhead is $O(mn|p|)$.

Similarly, the computation complexities of \textsf{Setup}, \textsf{Encrypt} and \textsf{Sum} in the sum protocol are $O(1)$, $O(1)$ and $O(m)$ respectively, and they are executed for only once in the scheme. Hence, the communication overhead of \textsf{Setup}, \textsf{Encrypt} and \textsf{Sum} are $O(|p^2|)$ bits, $O(|p^2|)$ bits and $O(m|p^2|)$ bits respectively ($|p^2|$ is the big length of $p^2$).

Note that $|p^2| = 2|p|$. Then, the total complexity of aggregator and participants are as follows:

\begin{table}[!h]
\centering \caption{One Aggregator Model}
\centering
\begin{tabular}{r|c|c}
\hline\hline
Aggregator & Computation & Communication (bits)\\
\hline
\textsf{Product} (Product) & $O(mn)$ & $O(mn|p|)$ \\
\textsf{Sum} (sum) & $O(m)$ & $O(m|p|)$ \\
\hline
Per Participant& Computation & Communication (bits)\\
\hline
\textsf{Setup} (Product) & $O(1)$ & $O(|p|)$ \\
\textsf{Encrypt} (Product) & $O(m)$ & $O(m|p|)$ \\
\textsf{Setup} (sum) &$O(1)$& $O(|p|)$ \\
\textsf{Encrypt} (sum) & $O(1)$ & $O(|p|)$ \\
\hline\hline
\end{tabular}
\end{table}

\subsubsection{Participants Only Model}

In the Participants Only Model, participants broadcast ciphertexts to others, and calculates the products and sums themselves, therefore the complexities are shown as below:

\begin{table}[!h]
\centering \caption{Participants Only Model}
\centering
\begin{tabular}{r|c|c}
\hline\hline
Per Participant  & Computation & Communication (bits)\\
\hline
\textsf{Setup} (Product) & $O(1)$ & $O(|p|)$ \\
\textsf{Encrypt} (Product) & $O(m)$ & $O(mn|p|)$ \\
\textsf{Product} (Product) & $O(mn)$ & $O(mn|p|)$ \\
\textsf{Setup} (sum) &$O(1)$& $O(|p|)$ \\
\textsf{Encrypt} (sum) & $O(1)$ & $O(m|p|)$ \\
\textsf{Sum} (sum) & $O(m)$ & $O(m|p|)$ \\
\hline
\hline
\end{tabular}
\end{table}

Note that the communication overhead is balanced in the Participants Only Model, but the system-wide communication overhead is increased a lot. In the One Aggregator Model, the system-wide communication overhead is:
\begin{displaymath}
O(mn|p|) + O(m|p|) + n\cdot  O(|p|)=O(mn|p|) ~~~~\text{(bits)}
\end{displaymath}

However, in the Participants Only Model, the system-wide communication complexity is:
\begin{displaymath}
n \cdot O(|p|) + n\cdot O(m|p|) + n\cdot O(mn|p|)=O(mn^2|p|)~~~~ \text{(bits)}
\end{displaymath}

\section{Performance Evaluation by Implementation}\label{section:implementation}
\label{sec:evalaution}


We conduct extensive evaluations of our protocols. Our simulation result  shows that the computation complexity of our protocol is indeed linear to the number of participants. To simulate and measure the computation overhead, we used GMP library to implement large number operations in our protocol in a computer with Intel i7-2620M @ 2.70GHz CPU and 2GB of RAM, and each result is the average time measured in the 100,000 times of executions. Also, the input data $x_i$ is of 20-bit length, the $q$ is of 256-bit length, and $p$ is roughly of 270-bit length. That is, $x_i$ is a number from $[0,2^{20}-1]$ and $q$ is a uniform random number chosen from $[0,2^{256}-1]$.

In this simulation, we measured the total overhead of our novel product protocol and sum protocol (the second sum protocol) proposed in the Section \ref{section:preliminaries}). Here, we measured the total computation time spent in calculating the final result of $n$ data (including encryption by $n$ participants and the decryption by the aggregator). Since we only measure the computation overhead, there is no difference between One Aggregator Model and Participants Only Model.

\begin{figure}[!th]
  \centering
  \begin{tabular}{cc}
  \includegraphics[width=0.22\textwidth]{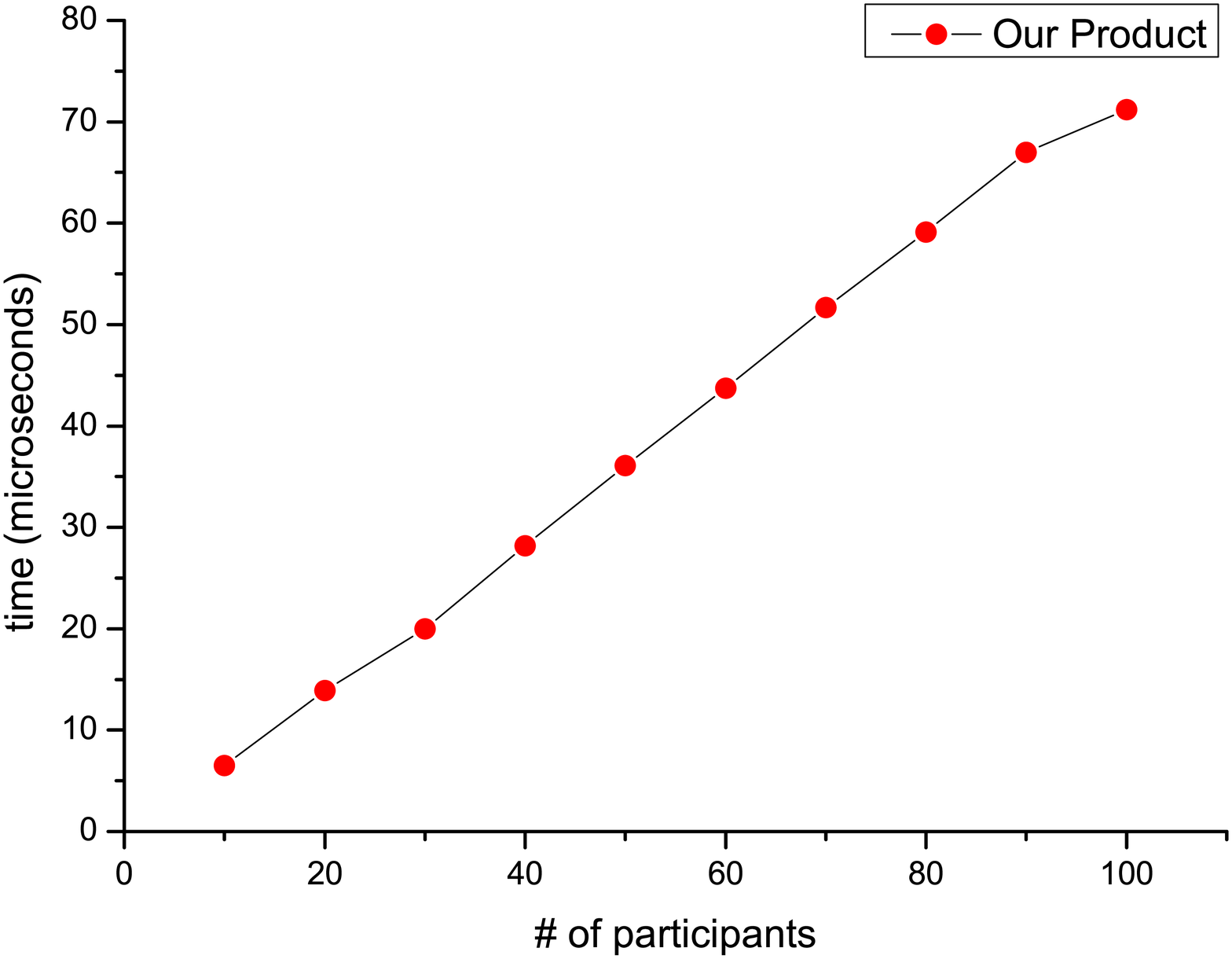} &
    \includegraphics[width=0.22\textwidth]{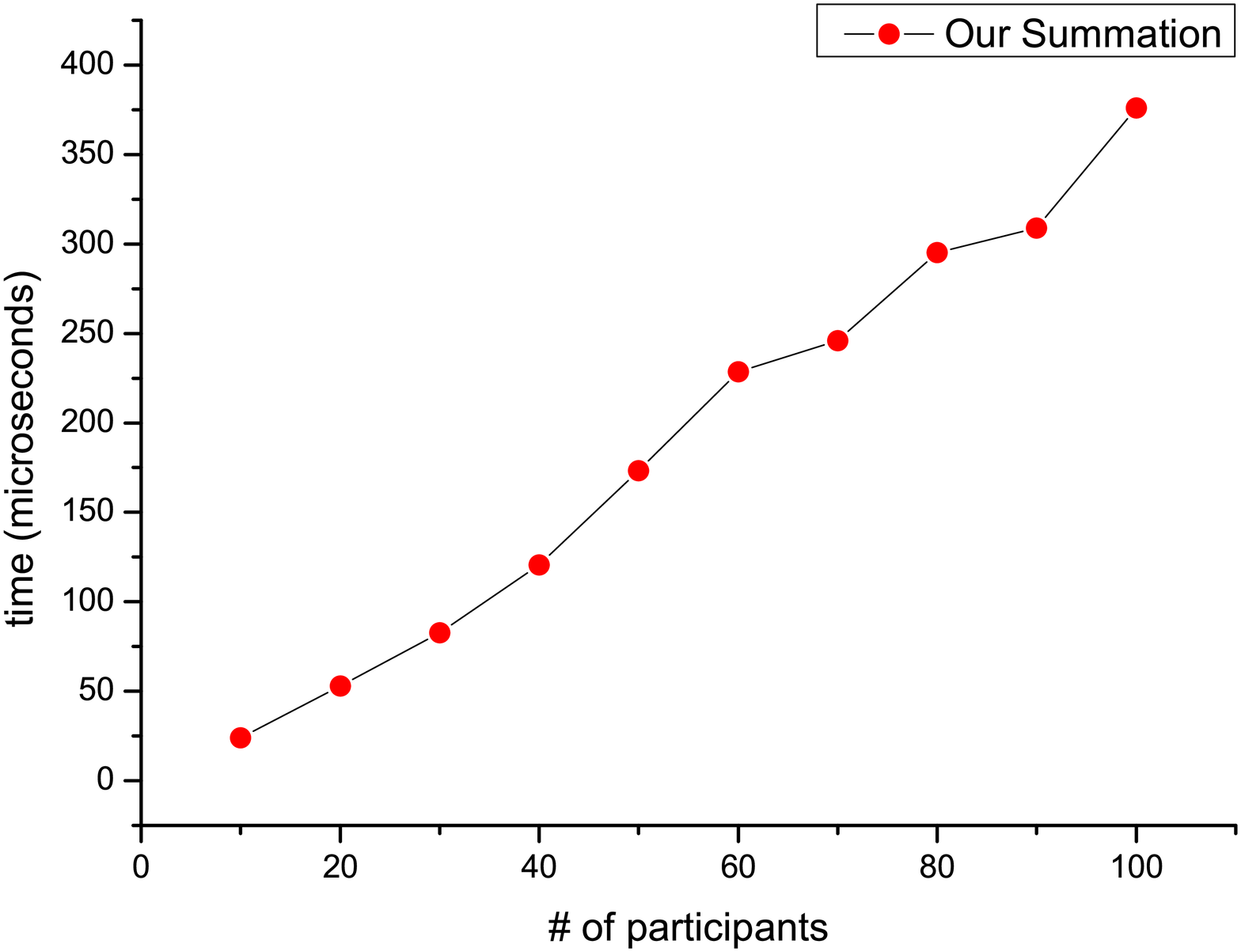} \\
(a) product & (b) sum
\end{tabular}
  \caption{Running time for product and sum calculation.}
  \label{fig:prod}
    \label{fig:sum}
\end{figure}

First of all, the computation overhead of each protocol is indeed proportional to the number of participants. Also, the sum protocol needs much more time. This is natural because parameters in the sum protocol are in $\zgroup_{p^2}$, which are twice of the parameters in the product protocol in big length (they are in $\zgroup_p$).

Multivariate polynomial evaluation is composed of $m$ products and one sum, so its computation overhead is barely the combination of the above two protocols' overhead.

We further compare the performance of our protocol with other existing multi party computation system implemented by Ben \textit{et al.} \cite{ben2008fairplaymp} (FairplayMP). They implemented the BMR protocol \cite{beaver1990round}, which requires constant number of communication rounds regardless of the function being computed. Their system provides a platform for general secure multi-party computation (SMC), where one can program their secure computation with Secure Function Definition Language (SFDL). The programs wrote in SFDL enable multiple parties to jointly evaluate an arbitrary sized boolean circuit. This boolean circuit is same as the garbled circuit proposed by Yao's 2 Party Computation (2PC) \cite{yao1982protocols}\cite{yao1986generate}.

In Ben's setting, where they used a grid of computers, each with two Intel Xeon 3GHz CPU and 4GB of RAM, they achieved the computation time as following tables when they have 5 participants:

\begin{table}[!h]
\centering \caption{Run time (milliseconds) of FairplayMP\cite{ben2008fairplaymp}}
\begin{tabular}{r|c|c|c|c|c|c}
\hline
Gates & 32 & 64 & 128 & 256 & 512 & 1024\\
\hline
Per Participant & 64  & 130 & 234 & 440 & 770 & 1394 \\
\hline
\end{tabular}
\end{table}

One addition of two $k$-bit numbers can be expressed with $k+1$ XOR gates and $k$ AND gates. Therefore, if we set the length of input data as 20 bits (which is approximately 1 million), we need 41 gates per addition in FairplayMP system. When we conduct 26 additions (which is equivalent to 1066 gates) in our system, the total computation time is 72.2 microseconds, which is $2\times 10^4$ times faster than the FairplayMP, which needs 1.394 seconds to evaluate a boolean circuit of 1024 gates. Even if we did not consider the aggregator's computation time in FairplayMP because they did not provide pure computation time (they provided the total run time including communication delay for the aggregator), our addition is already faster than their system. Obviously, the multiplication is much faster since it is roughly 8 times faster than the addition in our system.

We also compare our system with an efficient homomorphic encryption implementation \cite{lauter2011can}. Lauter \textit{et al.} proposed an efficient homomorphic encryption scheme which limits the total number of multiplications to a small number less than 100. If only one multiplication is allowed in their scheme (the fastest setting) and length of the modulus $q$ is 1024, it takes 1 millisecond to conduct an addition and 41 milliseconds to conduct a multiplication. In our system, under the same condition, it takes 16.2 microseconds to conduct an addition and 0.7 microseconds to conduct a multiplication, which are approximately 100 times and $6\times 10^4$ times faster respectively. They implemented the system in a computer with two Intel 2.1GHz CPU and 2GB of RAM. Even if considering our computer has a higher clock CPU, their scheme is still much slower than ours.

\begin{table}[!h]
\centering \caption{Comparison between \cite{lauter2011can} and our system}
\begin{tabular}{r|c|c}
\hline
& Addition & Multiplication \\
\hline
Lauter \cite{lauter2011can} & 1 millisecond & 41 milliseconds\\
\hline
Ours & 16.2 microseconds & 0.7 microseconds \\
\hline
\end{tabular}
\end{table}

The purpose of above two systems are quite different from ours, the first FairplayMP is for general multi-party computation and the second homomorphic encryption system is for general homomorphic encryption. They also provide a much higher level of security than ours since they achieve differential privacy, however, the comparison above does show the high speed of our system while our security level is still acceptable in real life applications, and this is one of the main contributions of this paper.

\section{Conclusion}
\label{sec:conclusion}

In this paper, we successfully achieve a privacy-preserving multivariate polynomial evaluation without secure communication channels by introducing our novel secure product and sum calculation protocol. We also show in the discussion that our proposed construction is efficient and secure enough to be applicable in real life.
However, our scheme discloses each product part in the polynomial, which gives unnecessary information to attackers. Therefore, our next research will be minimizing the information leakage during the computation and communication. 
Another future work is to design privacy preserving data releasing protocols such that certain functions can be evaluated correctly while certain functional privacy can be protected.


\begin{thebibliography}{10}
\providecommand{\url}[1]{#1}
\csname url@samestyle\endcsname
\providecommand{\newblock}{\relax}
\providecommand{\bibinfo}[2]{#2}
\providecommand{\BIBentrySTDinterwordspacing}{\spaceskip=0pt\relax}
\providecommand{\BIBentryALTinterwordstretchfactor}{4}
\providecommand{\BIBentryALTinterwordspacing}{\spaceskip=\fontdimen2\font plus
\BIBentryALTinterwordstretchfactor\fontdimen3\font minus
  \fontdimen4\font\relax}
\providecommand{\BIBforeignlanguage}[2]{{%
\expandafter\ifx\csname l@#1\endcsname\relax
\typeout{** WARNING: IEEEtranS.bst: No hyphenation pattern has been}%
\typeout{** loaded for the language `#1'. Using the pattern for}%
\typeout{** the default language instead.}%
\else
\language=\csname l@#1\endcsname
\fi
#2}}
\providecommand{\BIBdecl}{\relax}
\BIBdecl

\bibitem{beaver1990round}
D.~Beaver, S.~Micali, and P.~Rogaway, ``The round complexity of secure
  protocols,'' in \emph{Proceedings of the twenty-second annual ACM symposium
  on Theory of computing}.\hskip 1em plus 0.5em minus 0.4em\relax ACM, 1990,
  pp. 503--513.

\bibitem{ben2008fairplaymp}
A.~Ben-David, N.~Nisan, and B.~Pinkas, ``Fairplaymp: a system for secure
  multi-party computation,'' in \emph{Proceedings of the 15th ACM conference on
  Computer and communications security}.\hskip 1em plus 0.5em minus 0.4em\relax
  ACM, 2008, pp. 257--266.

\bibitem{bethencourt2007ciphertext}
J.~Bethencourt, A.~Sahai, and B.~Waters, ``Ciphertext-policy attribute-based
  encryption,'' in \emph{Security and Privacy, 2007. SP'07. IEEE Symposium
  on}.\hskip 1em plus 0.5em minus 0.4em\relax Ieee, 2007, pp. 321--334.

\bibitem{boneh1998decision}
D.~Boneh, ``The decision diffie-hellman problem,'' \emph{Algorithmic Number
  Theory}, pp. 48--63, 1998.

\bibitem{castelluccia2009efficient}
C.~Castelluccia, A.~Chan, E.~Mykletun, and G.~Tsudik, ``Efficient and provably
  secure aggregation of encrypted data in wireless sensor networks,'' \emph{ACM
  Transactions on Sensor Networks (TOSN)}, vol.~5, no.~3, p.~20, 2009.

\bibitem{castelluccia2005efficient}
C.~Castelluccia, E.~Mykletun, and G.~Tsudik, ``Efficient aggregation of
  encrypted data in wireless sensor networks,'' in \emph{Mobile and Ubiquitous
  Systems: Networking and Services, 2005. MobiQuitous 2005. The Second Annual
  International Conference on}.\hskip 1em plus 0.5em minus 0.4em\relax IEEE,
  2005, pp. 109--117.

\bibitem{clifton2002tools}
C.~Clifton, M.~Kantarcioglu, J.~Vaidya, X.~Lin, and M.~Zhu, ``Tools for privacy
  preserving distributed data mining,'' \emph{ACM SIGKDD Explorations
  Newsletter}, vol.~4, no.~2, pp. 28--34, 2002.

\bibitem{diffie1976new}
W.~Diffie and M.~Hellman, ``New directions in cryptography,'' \emph{Information
  Theory, IEEE Transactions on}, vol.~22, no.~6, pp. 644--654, 1976.

\bibitem{dong2011secure}
W.~Dong, V.~Dave, L.~Qiu, and Y.~Zhang, ``Secure friend discovery in mobile
  social networks,'' in \emph{INFOCOM, 2011 Proceedings IEEE}.\hskip 1em plus
  0.5em minus 0.4em\relax IEEE, 2011, pp. 1647--1655.

\bibitem{elgamal1985public}
T.~ElGamal, ``A public key cryptosystem and a signature scheme based on
  discrete logarithms,'' in \emph{Advances in Cryptology}.\hskip 1em plus 0.5em
  minus 0.4em\relax Springer, 1985, pp. 10--18.

\bibitem{gentry2009fully}
C.~Gentry, ``Fully homomorphic encryption using ideal lattices,'' in
  \emph{Proceedings of the 41st annual ACM symposium on Theory of
  computing}.\hskip 1em plus 0.5em minus 0.4em\relax ACM, 2009, pp. 169--178.

\bibitem{gentry2011implementing}
C.~Gentry and S.~Halevi, ``Implementing gentry’s fully-homomorphic encryption
  scheme,'' \emph{Advances in Cryptology--EUROCRYPT 2011}, pp. 129--148, 2011.

\bibitem{goldreich1998secure}
O.~Goldreich, ``Secure multi-party computation,'' \emph{Manuscript. Preliminary
  version}, 1998.

\bibitem{he2007pda}
W.~He, X.~Liu, H.~Nguyen, K.~Nahrstedt, and T.~Abdelzaher, ``Pda:
  Privacy-preserving data aggregation in wireless sensor networks,'' in
  \emph{INFOCOM 2007. 26th IEEE International Conference on Computer
  Communications. IEEE}.\hskip 1em plus 0.5em minus 0.4em\relax IEEE, 2007, pp.
  2045--2053.

\bibitem{jung2013cloud}
T.~Jung, X.~Li, Z.~Wan, and M.~Wan, ``Privacy preserving cloud data access with
  multi-authorities,'' in \emph{IEEE INFOCOM}, 2013.

\bibitem{lauter2011can}
K.~Lauter, M.~Naehrig, and V.~Vaikuntanathan, ``Can homomorphic encryption be
  practical,'' \emph{Preprint}, 2011.

\bibitem{li2013search}
X.~Li and T.~Jung, ``Search me if you can: privacy-preserving location query
  service,'' in \emph{IEEE INFOCOM}, 2013.

\bibitem{menezes1997handbook}
A.~Menezes, P.~Van~Oorschot, and S.~Vanstone, \emph{Handbook of applied
  cryptography}.\hskip 1em plus 0.5em minus 0.4em\relax CRC, 1997.

\bibitem{sheikh2009privacy}
R.~Sheikh, B.~Kumar, and D.~Mishra, ``Privacy preserving k secure sum
  protocol,'' \emph{Arxiv preprint arXiv:0912.0956}, 2009.

\bibitem{shi2011privacy}
E.~Shi, T.~Chan, E.~Rieffel, R.~Chow, and D.~Song, ``Privacy-preserving
  aggregation of time-series data,'' in \emph{Proceedings of NDSS}, vol.~17,
  2011.

\bibitem{verykios2004state}
V.~Verykios, E.~Bertino, I.~Fovino, L.~Provenza, Y.~Saygin, and Y.~Theodoridis,
  ``State-of-the-art in privacy preserving data mining,'' \emph{ACM Sigmod
  Record}, vol.~33, no.~1, pp. 50--57, 2004.

\bibitem{yao1982protocols}
A.~Yao, ``Protocols for secure computations,'' in \emph{Proceedings of the 23rd
  Annual Symposium on Foundations of Computer Science}, 1982, pp. 160--164.

\bibitem{yao1986generate}
------, ``How to generate and exchange secrets,'' in \emph{Foundations of
  Computer Science, 1986., 27th Annual Symposium on}.\hskip 1em plus 0.5em
  minus 0.4em\relax IEEE, 1986, pp. 162--167.

\bibitem{zhang2013verifiable}
L.~Zhang, X.~Li, Y.~Liu, and T.~Jung, ``Verifiable private multi-party
  computation: ranging and ranking,'' in \emph{IEEE INFOCOM Mini-Conference},
  2013.

\end{thebibliography}


\end{document}